\documentclass[12pt,tightenlines,prd]{revtex4}

\usepackage{amsmath}
\usepackage{latexsym}

\providecommand{\sgn}[1]{\mspace{2mu}\mathrm{sgn}\mspace{2mu}#1\,}
 \providecommand{\K}{\mathrm{K}}

\begin{document}






\title{Vacuum-polarization near cosmic string in
RS2 brane world}
\author{Yuri Grats}
\email{grats@string.phys.msu.su}
\author{Anton Rossikhin}
 \affiliation{M.V. Lomonosov Moscow State University,
119899, Moscow, Russia}

\begin{abstract}
Gravitational field of cosmic strings in theories with extra
spatial dimensions must differ significantly from that in the
Einstein's theory. This means that all gravity induced properties
of cosmic strings need to be revised too. Here we consider the
effect of vacuum polarization outside a straight infinitely thin
cosmic string embedded in a RS2 brane world. Perturbation
technique combined with the method of dimensional regularization
is used to calculate ${\langle T_{\mu}^{\nu}
 \rangle}_{vac}^{ren}$ for a massless scalar field.
\end{abstract}

\maketitle

\section{Introduction}

{
  Now it is well understood that topological defects in
quantum field theory may play an important role in physical
phenomena at very different spacetime scales. In particular,
vertex line defects describable on a macroscopic scale as cosmic
strings are most likely to be actually generated at phase
transitions in the early universe~\cite{ForEvCS}.  Recent
cosmological data, in particular power spectrum of the microwave
background, are not consistent with the density fluctuations
predicted by the simplest cosmic string scenario.  Nevertheless,
some combination of density fluctuations produced by strings and
inflation is possible. Moreover, gravitational properties of
cosmic strings become the question of increasing interest because
strings are predicted by many of the commonly considered models.

In a lot of applications radius of curvature of a string is much
larger than its transverse diameter.  So, string can be considered
to be infinitely thin, and one can use the so-called world sheet
approximation.  In this approximation four-dimensional Einstein's
gravity leads to the result that the spacetime of an infinitely
thin straight cosmic string is a direct product of the
two-dimensional Minkowski plain and a cone. Corresponding Riemann
tensor vanishes everywhere except on the world sheet of the
string, where it has a $\delta$-like singularity. So, in this
theory straight conical defects do not affect the local geometry
of the spacetime but change its global properties, and the effects
of conical structure on surrounding classical and quantized matter
fields are purely nonlocal (topological).  At the same time, it
seems unlikely that at the sufficiently large energy scales
Einstein's theory describes gravity accurately. This means that
alternative approaches, in particular those formulated in
spacetime of more than four spacetime dimensions, must be
considered too.

Until recently,  Kaluza-Klein theories with a small size of extra
dimensions have been of the most interest. At the present time,
however, the brane world picture becomes more and more popular. In
this approach our physical universe is considered to be a
$3$-brane embedded into a higher dimensional bulk (for the review
of this problem see~\cite{Rub} and references therein).  Thus,
gravity becomes multidimensional while ordinary matter fields are
still propagate in the world with three spatial dimensions.

Here we consider a so-called RS2 model with one infinite extra
dimension and matter localized on a single positive tension
brane~\cite{BWold}. The main goal of our paper is to demonstrate
once more that extra dimensions are not hidden, and there are
nontrivial effects of the higher-dimensional gravity on the fields
trapped to a brane.

This work is devoted to the  effect of vacuum polarization near
straight cosmic string imbedded in the brane of a RS2 world.  Our
investigation was stimulated by

\noindent {\bf i.}\quad\quad\!{}the paper~\cite{Aref00}, in which
a globally consistent expression for the linearized gravitational
field in the Randall-Sundrum background with a matter on the brane
is found;

\noindent {\bf ii.}\quad\,{}the paper~\cite{Davis00}, in which the
gravitational properties of a straight cosmic string in the RS2
brane world are investigated;

\noindent {\bf iii.}\quad{}our paper~\cite{Grats95}, in which  it
is shown that perturbative methods can be used in investigation of
vacuum polarization in the spacetime of a single string, and in
the spacetime of multiple cosmic strings.

We restrict our consideration by the case of a massless scalar
field, because it is obvious enough that for the fields of higher
spin the result will be the same up to the numerical coefficient.

The paper is organized as follows. We intend to use the method of
dimensional regularization. So, in Sec.\ref{LG}, the main results
of the paper~\cite{Aref00} are extended to the case of RS2 type
world of dimension $p+1+1$. Then we consider the particular case
when $p$-brane contains a multidimensional generalization of an
``ordinary'' straight string. Below we will use the term string
for this co-dimension two object.  In Sec.  \ref{SGF}, we
calculate the Euclidean Green's function for a massless scalar
field on the background under consideration.  In Sec.  \ref{EMT},
the method of dimensional regularization is used to obtain  the
expression for the renormalized vacuum averaged energy-momentum
tensor.  In Sec. \ref{Conclusion}, we summarize our results.

In this paper, we  use the metric with the signature
$\left(-+\cdots+\right)$ and the system of units $\hbar =c =1$.

}

\section{Gravitational field of a straight
 cosmic string in RS2
spacetime of dimension $p+1+1$} \label{LG}

Let us start by the formulation of a linearized gravity on a
$p$-brane imbedded in a bulk of dimension $p+1+1$. In this section
we briefly reproduce the main ideas of the paper~\cite{Aref00}.

The action under consideration reads
 \begin{equation}
 \label{action}
S =\frac1{16\pi G_{n+1}}\int d^n x \int dy \sqrt{-g} (R-2\Lambda)
+ \frac1{16\pi G_{n+1}}\int_{\text{brane}} d^n x \sqrt{-\hat g}
\left(\sigma +16\pi G_{n+1}\mathcal{L}_{\text{matter}}\right)\ ,
 \end{equation}
where $n = p+1$, $g_{ab}$ $(a,b=0,1,\ldots,n-1,y)$ is the
metric of the bulk space-time, $\hat g_{\mu\nu}$
$(\mu,\nu=0,1,\ldots,n-1)$ is the induced metric on the brane,
$G_{n+1}\equiv 1/M^{n-1}$ and $M$ are the bulk gravitational
constant and the bulk Plank mass correspondingly. We shell denote
coordinates on the $p$-brane as $x^{\mu}$, $y$ is coordinate
transverse to the brane.

To obtain  multidimensional solution of the Randall-Sundrun type
we must choose
\begin{equation}
\label{finetune} \Lambda = -\frac{n(n-1)}{2} k^2 \quad \text{and}
\quad \sigma = -4(n-1)k\ .
\end{equation}
In this case the metric
\begin{equation} \label{bmetric}
ds^2= e^{-2k|y|} \eta_{\mu\nu} dx^\mu dx^\nu +dy^2
\end{equation}
is the solutions for $\mathcal{L}_{\text{matter}}=0$.

As it was shown in the paper~\cite{Aref00}, there is a gauge, in
which the position of the 3-brane remains fixed at $y=0$ even when
matter is present. Proceeding along the same line, let us split up
the bulk metric tensor as
\begin{equation} \label{metric1}
  \left( g_{ab} \right) =
 \begin{pmatrix} \hat g_{\mu\nu} &N_\mu\\ N_\nu & N_\lambda N^\lambda +
   N^2 \end{pmatrix}, \qquad \left( g^{ab} \right) = \frac1{N^2}
 \begin{pmatrix} N^2 \hat g^{\mu\nu} + N^\mu N^\nu & - N^\mu\\ - N^\nu
   & 1 \end{pmatrix}, \end{equation}
where $N$ and $N_\mu$ are the lapse function and shift vector,
respectively.

For the linearization of the Einstein's equations, we write the
induced metric as
  \begin{equation} \label{ghat}
  \hat g_{\mu\nu} = e^{-2k|y|} \left( \eta_{\mu\nu}+
  \gamma_{\mu\nu} \right)\ ,
 \end{equation}
and consider $\gamma_{\mu\nu}$, $N_\mu$ and $\phi\equiv N^2-1$
to be small perturbations.

In this approximation the gauge conditions for the $p$-brane to
remain at $y=0$ have the form
  \begin{gather}
\label{i:nmuphi}
  N_\mu = - \frac{\sgn{y}}{2nk} \gamma_{,\mu}\ , \quad
  \phi = - \frac{\sgn{y}}{nk} \gamma_{,y}\ , \\
\label{i:nmuphi1} \tilde \gamma_{\mu\nu}^{,\mu}=0
\end{gather}
where $\gamma=\eta^{\mu\nu}\gamma_{\mu\nu}$ and $\tilde
\gamma_{\mu\nu}\equiv \gamma_{\mu\nu} - \frac1n \eta_{\mu\nu}
  \gamma$ is the traceless part of $\gamma_{\mu\nu}$.

Following~\cite{Aref00} we obtain that the resulting equation  for
the metric perturbation $ \tilde\gamma_{\mu\nu}$ is
\begin{multline}
\label{eqmot1a}
  \partial_y \left( e^{-2k|y|} \partial_y \tilde\gamma_{\mu\nu}
  \right) - (n-2)k\sgn{y} e^{-2k|y|} \partial_y \tilde\gamma_{\mu\nu}
  + \Box \tilde\gamma_{\mu\nu} \\
  = -16\pi G_{n+1} \delta(y) \left[ t_{\mu\nu}
  - \frac1{n-1} \left( \eta_{\mu\nu} - \frac{\partial_\nu
  \partial_\mu}{\Box}\right)t \right]\
\end{multline}
with a constrain
\begin{equation} \label{dirbcgamma}
\Box \gamma\Bigr|_{y=0}=\frac{8\pi G_{n+1} nk}{n-1}t\ .
\end{equation}
In the last two equations  $t_{\mu\nu}$ is the energy-momentum
tensor of a matter sitting on the brane and
$t=\eta^{\mu\nu}t_{\mu\nu}$.

Now, it is easy to show that for $k>0$ (Randall-Sundrum
case) and $q^2>0$ the final solution for $\tilde\gamma_{\mu\nu}$
in the momentum space reads
\begin{equation}
\label{solution}
  \tilde\gamma_{\mu\nu}(q,y) = \frac{8\pi  G_{n+1}}{|q|} \left[ t_{\mu\nu}
  - \frac1{n-1}\left( \eta_{\mu\nu} -\frac{q_\mu q_\nu}{q^2}
  \right) t\right] e^{nk|y|/2}
  \frac{\K_{n/2}\left(e^{k|y|} |q|/k \right)}{\K_{n/2-1}(|q|/k)}\ .
\end{equation}

{\sloppy

In $n = p+1$ spacetime dimensions the energy-momentum tensor
of a straight infinitely thin string (object of co-dimension two)
has the form
\begin{equation}
t_{\mu\nu}=-\mu\delta^2(\vec x)\left(\eta_{\mu\nu}-\sum_{i=1}^2
\delta_{i\mu}\delta_{i\nu}\right)\ ,
\end{equation}
where $\vec x=(x^1, x^2)$ and $\mu$ is the energy per unit volume
of the string.

For this particular case the traceless part of the metric on the brane
takes the form
\begin{equation}
\label{solution1}
  \tilde\gamma_{\mu\nu}(x,0) = {8\pi G_{n+1}\mu}
\int \frac{d^2\vec q}{(2\pi)^2}e^{i\vec q\vec x}
\frac{\K_{n/2}\left(q/k \right)}{q\K_{n/2-1}(q/k)} W_{\mu\nu}(\vec
q)
  \ ,
\end{equation}
where $\vec q=(q_1, q_2)$, $q = |\vec q|$ and
\begin{equation}
\label{sol1}
 W_{\mu\nu}(\vec q)=\frac1{n-1}\biggl[ (n-1)\sum_{i=1}^2
\delta_{i\mu}\delta_{i\nu}
  - \eta_{\mu\nu} -(n-2)\frac{q_\mu q_\nu}{q^2}
  \biggr] \ ,
\end{equation}
with $q^{\mu}=(0,q^1, q^2, 0, \ldots, 0)$. \noindent As for the
trace of the metric, it can be calculated using
Eq.~(\ref{dirbcgamma})
\begin{equation}
\label{solution2}
 \gamma(x)=8\pi k G_{n+1}\mu\frac{n(n-2)}{n-1}\int
 \frac{d^2\vec q}{(2\pi)^2}\frac{e^{i\vec q\vec x}}{q^2}\ .
\end{equation}

The solution obtained~(\ref{solution1}) can be written in  the form
$\tilde\gamma_{\mu\nu}(x,0)=\tilde\gamma^0_{\mu\nu}(x)
+\tilde\gamma^1_{\mu\nu}(x)$,
\begin{gather}
\label{solution3}
 \tilde\gamma^0_{\mu\nu}(x)= {8\pi G_{n+1}(n-2)k\mu}
\int \frac{d^2\vec q}{(2\pi)^2}e^{i\vec q\vec x}
\frac{1}{q^2}W_{\mu\nu}(\vec q)\ , \\ \label{solution4}
 \tilde\gamma_{\mu\nu}^1 (x) = {8\pi G_{n+1}\mu}
\int \frac{d^2\vec q}{(2\pi)^2}e^{i\vec q\vec x}
\frac{\K_{n/2-2}\left(q/k \right)}{q\K_{n/2-1}(q/k)}
W_{\mu\nu}(\vec q)\ ,
\end{gather}
where $\tilde\gamma^0_{\mu\nu}$ is the solution corresponding to
the case of a string in the $n$-dimensional Einstein's gravity, while
the second term is the contribution from the bulk.

Now we can obtain the expressions for the intrinsic
Ricci tensor
\begin{equation}
\label{cur1} {\hat R}_{\mu\nu}= 8\pi G_{n+1}k\mu\frac{n-2}{2}
\delta^2(\vec x) \sum_{i=1}^2 \delta_{i\mu}\delta_{i\nu}
-\frac12\Box\tilde\gamma^1_{\mu\nu}(x)\ .
\end{equation}
and for the scalar curvature
\begin{equation}
\label{cur2} {\hat R}=16\pi G_{n+1}k\mu\frac{n-2}{2} \delta^2(\vec
x)
 \ .
\end{equation}

We see that, in contrast to its four-dimensional equivalent, the
spacetime of a string on a brane is not Ricci flat, but the
(linearized) intrinsic scalar curvature is still equal to zero
everywhere outside the core.

{\sloppy
\section{Euclidean Green's function}
\label{SGF}

Euclidean Green's function of a scalar field in a spacetime of the
dimension $n$ is the solution of the Poisson equation
\begin{equation}
\label{q1.5} \Bigl({\hat g^{\mu\nu}}_E{\hat \nabla}_{\mu}{\hat
 \nabla}_{\nu}-\xi \hat R\Bigr)G^E(x,x')=-
 \frac{\delta^n(x-x')}{\sqrt{{\hat g}_E (x)}}\ ,
\end{equation}
where ${\hat g_{\mu\nu}^E}$ is obtained from the metric under
consideration by the Wick rotation.

Proceeding along the same line as in the paper~\cite{Grats95}, let
us write this equation in the form
\begin{equation}
\label{new1} \Delta^n_0G^E(x,x')=-\delta^n(x-x')-VG^E(x-x')\ ,
\end{equation}
where
$\Delta^n_0\equiv\delta^{\mu\nu}\partial_{\mu}\partial_{\nu}$ is
the Laplase operator in the $n$-dimensional Euclidean space, and
\begin{equation}
V=\partial_{\mu}\left(\sqrt{\hat g}_E {\hat g}^{\mu\nu}_E
\partial_{\nu}\right)-\delta^{\mu\nu}\partial_{\mu}\partial_{\nu} -
\xi \hat R\ .
\end{equation}

When the use of perturbation theory is justified, the solution of
the equation~(\ref{new1}) can be written as
\begin{equation}
\label{Gr} G^E=G_0^E+G_0^EVG_0^E+G_0^EVG_0^EVG_0^E+\ldots,
\end{equation}
where $G_0^E$ is the Green's function of the Poisson equation  in
the $n$-dimensional Euclidean space.

}

If we restrict ourselves by the first-order correction to the
Green's function, we can represent the perturbation operator $V$
in the form
\begin{equation}
\label{V} V=-\tilde\gamma^{\mu\nu}\partial_{\mu}\partial_{\nu}+
\frac{n-2}{2n}\delta^{\mu\nu}\left(\partial_{\mu}
\gamma\partial_{\nu}+\gamma\partial_{\mu}\partial_{\nu}
\right)-\xi\hat R\ .
\end{equation}
In the last equation $\hat R$ is the first-order correction to the
intrinsic scalar curvature, and the explicit form of the
linearized metric, obtained in the previous section, was taken
into account.

As it was shown in the previous section,
(see~(\ref{cur2})), $\hat R$ is equal to zero outside the world
sheet of a string, where it has a $\delta$-like singularity.
Eq.(~\ref{q1.5}) is ill-defined for such a potential .  The
possible way to avoid this problem is to smooth out the
singularity~\cite{AlOt90,ALO92,FUSOL95}.  As the result, one finds
that when (at the end of calculations) the regularization of the
singularity is removed, the dependence of the Green's functions on
$\xi$ vanishes. So, this term can be ruled out of $V$.

Now using equations~(\ref{Gr}), (\ref{V})  and the explicit
expression for the function $G_0^E$, we find that the first-order
correction to the Euclidean Green's function $G_1^E=G_0^EVG_0^E$
reads
\begin{equation}
\label{SG} G_1^E=G_{Eins}+G_{bulk}\ ,
\end{equation}
where
\begin{equation}
\label{GEinst} G_{Eins}= -{8\pi G_{n+1} k
\mu}{(n-2)}\int\frac{d^2\vec q}{(2\pi)^2}\frac{e^{i\vec q \vec
x}}{q^2}\int
\frac{d^np}{(2\pi)^n}\frac{e^{-ip(x-x')}}{p^2(p-q)^2}\sum_{i=3}^n
p_i^2\
\end{equation}
coincides with the the expression obtained for the first-order
corrections to the Green's function  in the $n$-dimensional
locally flat conical space~\cite{Grats95}, while
\begin{equation}
\label{Gbulk} G_{bulk}=8\pi G_{n+1}\mu \int\frac{d^2\vec
q}{(2\pi)^2}e^{i\vec q\vec x} \frac{\K_{n/2-2}\left(q/k
\right)}{q\K_{n/2-1}\left(q/k\right)}\int
\frac{d^np}{(2\pi)^n}\frac{e^{-ip(x-x')}}{p^2(p-q)^2}W_{\mu\nu}(\vec
q) p^{\mu}p^{\nu}\
\end{equation}
is the contribution from the bulk.

\section{Calculation of the vacuum
expectation value \protect$\langle
T_{\mu}^{\nu}\rangle^{ren}_{vac}$ } \label{EMT}

To calculate the vacuum expectation value ${\langle
T_{\mu}^{\nu}\rangle}^{ren}_{vac}$, let us start from the formal
expression
\begin{equation} \label{T} \langle
T^{\nu}_{\mu}\rangle_{vac}=\lim_{x'\to
x}D^{\nu'}_{\mu}(x,x')G^E(x, x')\ ,
\end{equation}
where $D^{\nu'}_{\mu}$ is a differential operator whose explicit
form is determined by the classical expression for the
stress-energy tensor. For a massless scalar field with arbitrary
coupling and in the lowest order with respect to $\mu$, we have
\begin{equation}
D^{\nu'}_{\mu}=(1-2\xi)\nabla^{\nu'}_{\mu}-
2\xi\nabla^{\nu}_{\mu}+\left(2\xi-\frac12\right)
\delta^{\nu}_{\mu}\nabla^{\lambda'}_{\lambda}\ .
\end{equation}
Substituting the expression for $G_1^E$  into~(\ref{T}), we
obtain, that in our approximation the vacuum expectation value of
the energy-momentum tensor consists of two parts $${\langle
T_{\mu}^{\nu}\rangle}_{vac}= {\langle
T_{\mu}^{\nu}\rangle}_{Einst}+{\langle
T_{\mu}^{\nu}\rangle}_{bulk}\ ,$$ where
\begin{gather}
\label{TEinst} {\langle T_{\mu}^{\nu}\rangle}_{Einst}=-8\pi
G_{n+1}(n-2)k \mu\int\frac{d^2\vec q}{(2\pi)^2} \frac{e^{i\vec
q\vec x}}{q^2} \int
\frac{d^np}{(2\pi)^n}\frac{D_{\mu}^{\nu}(p,q)}{p^2(p-q)^2}{\sum_{i=3}^n
p_i^2}\ ,\\
 D_{\mu}^{\nu}(p,q)=p_{\mu}p^{\nu}-(1+2\xi)q_{\mu}p^{\nu}+2\xi
q_{\mu}q^{\nu}- \left(\xi-\frac14\right)q\delta_{\mu}^{\nu}\ ,
\end{gather}
and
\begin{equation}
\label{Tbulk} {\langle T_{\mu}^{\nu}\rangle}_{bulk}=8\pi
G_{n+1}\mu \int\frac{d^2\vec q}{(2\pi)^2}e^{i\vec q\vec x}
\frac{\K_{n/2-2}\left(q/k
\right)}{q\K_{n/2-1}\left(q/k\right)}\int \frac{d^np}{(2\pi)^n}
\frac{D_{\mu}^{\nu}(p, q)}{p^2(p-q)^2}
W_{\alpha\beta}p^{\alpha}p^{\beta}\ .
\end{equation}

Subsequent calculations can be performed using the method of
dimensional regularization along the same line as in the case of
the four-dimensional Einstein's theory. One must perform the
replacement $n\to D= n-2\epsilon$ and multiply the regularized
vacuum averaged stress-energy tensor by $(\lambda)^{n-D}$ (
$\lambda$ is an arbitrary parameter with the dimension of mass) to
restore the dimension of $\langle T_{\mu}^{\nu}\rangle_{vac}$.
After the regularization all the integrations in the
Eq.~(\ref{TEinst}) can be performed explicitly. Corresponding
calculations can be found in the paper~\cite{Grats95}. Using the
results of this paper we can write that for the  case of $3$-brane
$(n=4)$ the zero-zero component reads
\begin{equation}
\label{new5} {\langle
T_0^0\rangle}_{Einst}^{ren}=\frac{G_4\mu}{90{\pi}^2 r^4}+ (\xi -
\frac{1}{6})\frac{4G_4 \mu}{{\pi}^2 r^4}\ .
\end{equation}

Regularized contribution from the bulk has the form
\begin{multline}
\label{T1} {\langle T_{\mu}^{\nu}\rangle}^{reg}_{bulk}=\frac{8\pi
G_{n+1} k \mu}{4(D+1)(D/2-1)}
\Gamma(2-D/2)\frac{\Gamma^2(D/2)}{\Gamma(D)}\times\\ \times
\frac1{ (4\pi)^{D/2}} \int\frac{d^2\vec q}{(2\pi)^2}e^{i\vec q\vec
x} \frac{\K_{D/2-2}\left(q/k
\right)}{qk\K_{D/2-1}(q/k)}q^n\left(\frac{q}{\lambda}\right)^{D-n}
W_{\mu}^{\nu}(\vec q)\ .
\end{multline}

It is well known~\cite{Bunch79}, that in the smooth region of a
spacetime
\begin{multline}
\label{Tdiv} \langle
T_{\mu}^{\nu}\rangle_{div}=-\frac{1}{(4\pi)^{D/2}}
\left[\frac1{D-4}+\frac12\left\{C+\ln\left(\frac{{\lambda_0}^2}
{{\lambda}^2}\right)\right\} \right]\times \\ \times
\left(\frac1{90}H_{\mu}^{\nu}-\frac1{90}{}^{(2)}H_{\mu}^{\nu}+
\left(\frac16-\xi\right)^2{}^{(1)}H_{\mu}^{\nu}\right) \ .
\end{multline}
In the last expression  $\lambda_0$ is the infrared cutoff which
must be introduced in the case of massless fields, and
for our metric been taken in the linear approximation we find
\begin{gather}
H_{\mu\nu}=-4\Delta^n_0 \hat R_{\mu\nu}+2\hat R_{;\,\mu\nu}\\
{}^{(1)}H_{\mu\nu}=2\hat R_{;\,\mu\nu}-2\delta_{\mu\nu}\Delta^n_0
\hat R \\ {}^{(2)}H_{\mu\nu}=\hat
R_{;\,\mu\nu}-\frac12\delta_{\mu\nu}\Delta^n_0 \hat R-\Delta^n_0
\hat R_{\mu\nu} \ .
\end{gather}

As we know, outside the core of the string $\hat R=0$ and $\hat
R_{\mu\nu}=-\Delta^n_0\tilde\gamma^1_{\mu\mu}/2$. Combining all
the results above, for the case of $3$-brane we get
\begin{multline}
\label{Tren} {\langle
T_{\mu}^{\nu}\rangle}^{ren}_{bulk}=-\frac{1}{(4\pi)^{2}}
\frac{4\pi G_4\mu}{90}\int\frac{d^2\vec q}{(2\pi)^2}e^{i\vec q\vec
x} \frac{\K_{0}\left(q/k
\right)}{qk\K_{1}(q/k)}q^4\ln\left(\frac{q}{\lambda}\right)\times
\\ \times \biggl[ 3\sum_{i=1}^2
\delta_{i\mu}\delta_{i}^{\nu}
  - \delta_{\mu}^{\nu} -2\frac{q_\mu q^\nu}{q^2}
  \biggr]\ .
\end{multline}

From Eq~(\ref{Tren}) we can obtain that at large distances, when
$r\gg k^{-1}$, the contribution from the bulk is negligible
compared to ~(\ref{new5})
\begin{equation}
\langle T_0^0\rangle^{ren}_{bulk}=\frac1{(4\pi)^2}\frac{64
G_4\mu}{45}\frac1{k^2r^6}\left[\ln 2+3-C-2\ln (r\lambda) \right]\
,
\end{equation}
while near the core of the string, at $r\ll k^{-1}$,
\begin{equation}
\langle T_0^0\rangle^{ren}_{bulk}=\frac1{(4\pi)^2}
\frac{G_5\mu}{5}\frac1{r^5}\left[-\ln{2}+\frac{8}{3}-C -\ln
(r\lambda) \right]\ .
\end{equation}

We see, that contrary to the case of the Einstein's gravity the
expression for the total vacuum averaged $\langle
T_{\mu\nu}\rangle^{ren}_{vac}$ contains the $\lambda$-dependent
term which appears due to the renormalization procedure.

\section{Conclusions}
\label{Conclusion} In this paper we consider the vacuum
polarization effects outside a straight cosmic string in a RS2
brane world. It is shown, that in the  lowest order of the
perturbation theory $\langle T_{\mu\nu}\rangle^{ren}_{vac}$
consists of two parts. The first term depends on the angular
deficit only and reproduces the result obtained previously in the
Einstein's gravity.  This contribution is uniquely defined because
of the local flatness of the cosmic string spacetime in four
dimensions. The second contribution depends on the scale $k$ and
arises due to the possibility of gravity to propagate in the
five-dimensional bulk. Relative contribution of this term tends to
zero as $kr\to\infty$, but it becomes significant at short
distances. Another interesting feature is the appearance of an
arbitrary mass scale $\lambda$ in this part of $\langle
T_{\mu\nu}\rangle^{ren}_{vac}$. From the first glance the loss of
uniqueness must lead to some problems because the ordinary
approach to the back reaction problem consists of solving the
semiclassical Einstein's equations with $\langle
T_{\mu\nu}\rangle^{ren}_{vac}$ in its right-hand side. Analogous
question was considered in the case of a global monopole in the
four-dimensional Einstein's theory~\cite{Mazz91}, and as in the
case of the global monopole, it can be shown that any variation of
$\lambda$ may be absorbed by the coefficients before the
fourth-order terms appearing in the left-hand side of the one-loop
Einstein's equations.

\section{Acknowledgments}
This work was supported by the Russian Foundation for Basic
Research, grant 99-02-16132.

\end{document}